\documentclass[prd,preprint,showpacs,superscriptaddress,showkeys,floatfix,nofootinbib]{revtex4}
\pdfoutput=1
\usepackage{graphicx}
\usepackage{dcolumn}
\usepackage{bm}
\usepackage{color}
\usepackage{url}
\usepackage{amsmath}
\usepackage{amssymb}
\usepackage{mathrsfs}
\usepackage{accents}
\usepackage{epsfig}
\usepackage[lofdepth,lotdepth,caption=false]{subfig}
\usepackage[11pt]{moresize}
\usepackage{mathtools}
\usepackage{mathrsfs}
\newlength\figureheight 
\newlength\figurewidth 

\begin{document}

\title{Discrete symmetries and mixing of Dirac neutrinos}
\author{Arman Esmaili}
\affiliation{INFN, Laboratori Nazionali del Gran Sasso, Assergi (AQ), Italy}
\author{Alexei Yu.~Smirnov}
\affiliation{Max-Planck Institute for Nuclear Physics, Saupfercheckweg 1, D-69117 Heidelberg, Germany}

\date{\today} 

\begin{abstract}
We study mixing of the Dirac neutrinos in the residual symmetries approach. The key difference from the Majorana case is that the Dirac mass matrix may have larger symmetries: $G_\nu=\mathbf{Z}_{n}$ with $n\geq3$. The symmetry group relations have been generalized to the case of Dirac neutrinos. Using them we have found all new relations between mixing parameters and corresponding symmetry assignments which are in agreement with the present data. The viable relations exist only for the charged lepton residual symmetry $G_{\ell} = \mathbf{Z}_{2}$. The relations involve elements of the rows of the PMNS matrix and lead to precise predictions of the 2-3 mixing angle and certain ranges of the CP violation phase. For larger symmetries $G_{\ell}$, an agreement with data can be achieved if $\sim10\%$ corrections related to breaking of $G_{\ell}$ and $G_\nu$ are included. 
\end{abstract}


\maketitle

\section{Introduction}
\label{sec:intro}

The residual symmetry approach~\cite{Lam:2008sh} is one of the appealing possibilities to explain the lepton mixing. According to this approach, the mixing originates from different ways of the flavor symmetry breaking in the neutrino and charged lepton Yukawa interactions. These different ways of breaking lead to different residual symmetries, $G_\nu$ and $G_\ell$, of the neutrino and charged lepton mass matrices~\cite{Altarelli:2010gt,King:2013eh,King:2014nza}. The residual symmetries ensure certain form of the mass matrices, and consequently, the mixing matrix. The observed pattern of lepton mixing close to Tri-BiMaximal (TBM) mixing is hardly connected to the neutrino and charged lepton masses or ratios of masses. For this reason the ``generic'' symmetries $G_\nu$ and $G_\ell$ were used which exist for arbitrary values of masses (eigenvalues of the mass matrices). In this way symmetry provides complete control over the mixing. For the Majorana neutrinos maximal generic symmetry is given by the Klein group $\mathbf{Z}_2 \times \mathbf{Z}_2$. Depending on selected residual groups and symmetry assignments for the leptons, different mixing matrices and relations between the mixing matrix elements can be obtained. The general relations~\cite{Hernandez:2012sk,Hernandez:2012ra}, the case of $\mathbf{Z}_2$ symmetry~\cite{Ge} and maximal CP violation~\cite{cpmax} have been discussed. 

Unfortunately, realizations of this approach in consistent gauge models are rather complicated and not very convincing: apart from set of new fields (flavons), they contain a number of assumptions, new parameters and additional symmetries with \textit{ad hoc} charge assignments (see~\cite{Altarelli:2010gt,King:2013eh,King:2014nza,Smirnov:2011jv} for reviews). In view of this, existence of symmetries behind lepton mixing, and in particular, residual symmetries is still an open issue. 

The standard way to obtain predictions for mixing angles consists of model-building, construction of mass matrices, and finally, diagonalization of these matrices. It was realized, however, that predictions for mixing angles can be obtained from knowledge of the residual symmetries immediately without model-building~\cite{Hernandez:2012sk,Hernandez:2012ra}. The \textit{symmetry group relation} has been derived which includes the mixing matrix, $U_{\rm PMNS}$, and the matrices of transformations of the residual symmetries of neutrinos, $S$, and charged leptons, $T$, in mass basis~\cite{Hernandez:2012ra}. The symmetry group relation is an efficient tool to explore possible consequences of various symmetries. Once viable relations and corresponding symmetry assignments are realized, one can proceed with model-building.  

It is widely believed that neutrinos are Majorana particles and smallness of neutrino mass is related somehow to the Majorana nature of neutrinos. Therefore, most of the studies of discrete residual symmetries have been performed for Majorana neutrinos. However, it is possible that neutrinos are the Dirac particles and, in fact, there is a number of mechanisms and models which lead to small Dirac neutrino masses~\cite{modelDirac}, \textit{e.g.} Dirac seesaw, Peccei-Quinn symmetry, extra dimensional mechanisms, chiral symmetry, {\it etc}. Consequences of some discrete flavor symmetries for the Dirac neutrinos have been studied~\cite{fDirac}. 

In this paper we will consider applications of the residual symmetry approach to the Dirac neutrinos. For this we will use generalizations of the symmetry group relation. The goal is to see if new relations between the mixing parameters can be obtained in the case of Dirac neutrinos with the hope that their simpler realizations are possible in consistent gauge models.

The paper is organized as follows. In sec.~\ref{sec:residual} we consider the symmetry group relation for Dirac neutrinos. We study the case of $\mathbf{Z}_2$ residual symmetry for charged leptons ($m = 2$) in sec.~\ref{sec:cons}. $\mathbf{Z}_m$ symmetry with $m > 2$ for charged leptons will be explored in sec.~\ref{sec:m>2}. In sec.~\ref{sec:m3} we discuss some further considerations and generalizations. Discussion and conclusions are presented in sec.~\ref{sec:conc}.

\section{Symmetry group relations for Dirac neutrinos}
\label{sec:residual}

We introduce the diagonal mass matrices of charged leptons, $m_\ell$, and Dirac neutrinos, $m_\nu$, as well as the PMNS maxing matrix, $U_{\rm PMNS}$, according to the following Lagrangian
\begin{equation}
\mathcal{L} = \frac{g}{\sqrt{2}} \bar{\ell}_L U_{\rm PMNS} 
\gamma^\mu \nu_L W^+_\mu + \bar{E}_R m_\ell \ell_L + \bar{N}_R m_\nu \nu_L + {\rm h.c.}~.
\end{equation}
Here $\ell_L\!=\!(e_L,\mu_L,\tau_L)^T$, $E_R\!=\!(e_R,\mu_R,\tau_R)^T$, $\nu_L\!=\!(\nu_{1L},\nu_{2L},\nu_{3L})^T$ and $N_R=(\nu_{1R},\nu_{2R},\nu_{3R})^T$. Let $S$ and $T$ be the matrices of transformation of the left-handed (as well as right-handed) components of neutrinos and charged leptons that leave the mass matrices invariant:
\begin{equation} 
S^\dagger m_\nu S =  m_\nu \quad , \quad T^\dagger m_\ell T =  m_\ell~.
\end{equation}
So, $S$ and $T$ are generators of the residual symmetry groups $G_\nu$ and $G_\ell$ in the mass basis. For discrete finite symmetry groups, there should be integers $n$ and $m$ such that 
\begin{equation}\label{eq:finite}
S^n=\mathbb{I} \quad {\rm and} \quad T^m=\mathbb{I}~.
\end{equation}
Then the symmetry group condition reads~\cite{Hernandez:2012ra} 
\begin{equation}\label{eq:st} 
\left[ U_{\rm PMNS}\, S \, U_{\rm PMNS}^\dagger \, T \right]^p = \mathbb{I}~,
\end{equation}
where $p$ is an integer number. Equivalently the condition can be transformed to~\cite{Hernandez:2012ra}  
\begin{equation}\label{eq:grouprel}
{\rm Tr}\left[U_{\rm PMNS}\,S\,U_{\rm PMNS}^\dagger \,T\right]= a_p~,
\end{equation}
where
\begin{equation}\label{eq:sum}
a_p =\sum_{\beta=1}^{3} \lambda_\beta \quad , \quad \left(\lambda_\beta\right)^p = 1~, 
\end{equation}
{\it i.e.}, $\lambda_\beta$ are the $p$-roots of unity. The three selected roots in Eq.~(\ref{eq:sum}) satisfy the condition 
\begin{equation}\label{eq:prod}
\prod_{\beta=1}^{3} \lambda_\beta = 1~,
\end{equation}
which guarantees that the symmetry group $G_\nu\times G_\ell$ can be embedded in $SU(3)$. Values of $a_p$, for different $p$, which satisfy Eqs.~(\ref{eq:sum}) and (\ref{eq:prod}) are shown in Table~\ref{tab:ap}. Notice that for $p\leq3$ the value of $a_p$ is unique.  

\begin{table}[htdp]
\caption{Values of $a_p$ for different $p$.}
\begin{center}
\begin{tabular}{|c|c|}
\hline
$a_2$ & -1 \\
\hline
$a_3$ & 0 \\
\hline
$a_4$ & 1 , $-1\pm2i$ \\
\hline
$a_5$ & \begin{tabular}[c]{@{}c@{}} 1 , $\frac{1\pm\sqrt{5}}{2}$, \\ $\frac{-3-\sqrt{5}}{4} + i \frac{\sqrt{5(5+\sqrt{5})}}{2\sqrt{2}}$, \\ $\frac{-3+\sqrt{5}}{4} - i \frac{\sqrt{5(5-\sqrt{5})}}{2\sqrt{2}}$ \\ \end{tabular} \\
\hline
\end{tabular}
\end{center}
\label{tab:ap}
\end{table}

The relation in Eq.~(\ref{eq:grouprel}) is general and valid for both Majorana and Dirac neutrinos. Its derivation is completely the same in both cases.

In general, the symmetry transformation matrix $T$ can be written as
\begin{equation}
\label{eq:T}
T= {\rm diag} \left( e^{i\phi_e} , e^{i\phi_\mu} , e^{i\phi_\tau}  \right)~.
\end{equation}
The finiteness of the group, Eq.~(\ref{eq:finite}), allows to parametrize $\phi_\alpha$ as 
\begin{equation}
\phi_\alpha= \frac{2\pi k_\alpha}{m}~, \quad {\rm where} \quad 0\leq k_\alpha < m~,
\end{equation}
and $k_\alpha$ are integers. The condition $\det[T]=1$, which means that the symmetry group generated by $T$ can be embedded in $SU(3)$, implies that $k_e+ k_\mu+ k_\tau =  mq_k $, where $q_k = 1, 2$, and so for a given $q_k$ just two of $k_\alpha$ are independent. If one of $k_\alpha$ is zero, each transformation is determined by a single parameter $k$. For example, for $k_e=0$, we have 
$$T_{(e)} = {\rm diag} \left(1,e^{2\pi i k/m},e^{2\pi i (m-k)/m}\right)~,$$
where we introduced the subscript of $T$ which correspond to the lepton which does not transform under $T$. In the same way we can introduce independent symmetry transformations
$$T_{(\mu)} = {\rm diag} \left(e^{2\pi i k/m},1,e^{2\pi i (m-k)/m}\right) 
\quad , \quad T_{(\tau)} = {\rm diag} \left(e^{2\pi i k/m},e^{2\pi i (m-k)/m},1\right)~.$$
In general $m$ and $k$ in $T_{(\mu)}$ and $T_{(\tau)}$ can be different, so that the total symmetry group is $\mathbf{Z}_m \times \mathbf{Z}_{m'} \times \mathbf{Z}_{m''}$. We will keep the general form of $T$, as in Eq.~(\ref{eq:T}), for our calculations.

Similarly to (\ref{eq:T}), for the Dirac neutrinos, the transformation $S$ is given by
\begin{equation}\label{eq:S}
S= {\rm diag} \left( e^{i\psi_1}, e^{i\psi_2}, e^{i\psi_3} \right)~.
\end{equation}
From Eq.~(\ref{eq:finite}), we can parametrize the phases as $\psi_i=2\pi l_i/n$ with $0\leq l_i<n$. The condition $\det[S]=1$ leads to 
\begin{equation}\label{eq:det1}
\sum_{j = 1}^3 l_j = n q_l~, ~~~~ q_l = 1, 2~,
\end{equation}
so that two $l_i$ are independent. 

In general three independent generators can be introduced and the total neutrino mass matrix symmetry is $\mathbf{Z}_n \times \mathbf{Z}_{n'} \times \mathbf{Z}_{n''}$. For Majorana neutrinos the generic (mass independent) symmetry exists only for $n = 2$. So, the new possibilities which are specific to the Dirac neutrinos consist of symmetries with $n>2$. 

Recall that the symmetry group condition stems from the fact that $S$ and $T$ originate from the same finite discrete group. In the flavor basis the charged currents are diagonal and whole information on mixing is encoded in the mass matrix of neutrinos, which is now non-diagonal: $m_\nu^f =  U_{\rm PMNS} m_\nu U_{\rm PMNS}^\dagger$. In this basis the symmetry transformation is given by $S_U = U_{\rm PMNS} S U_{\rm PMNS}^\dagger$. Then, the condition that $S_U$ and $T$ form the same finite group is that their product belongs to the same group and some integer $p$ exist such that $\left( S_{U} T \right)^p = \mathbb{I}$. This condition coincides with the symmetry group relation in Eq.~(\ref{eq:st}). 

The parameters $n$, $m$ and $p$ in Eqs.~(\ref{eq:finite}) and (\ref{eq:st}) define the von Dyck group $D(n,m,p)$. The case of Majorana neutrinos discussed in~\cite{Hernandez:2012ra} is the special case with $n=2$. The von Dyck group is a finite group if 
\begin{equation}\label{eq:vondyck}
\frac{1}{n} + \frac{1}{m} + \frac{1}{p} > 1~.
\end{equation} 

Introducing the real and imaginary parts, $a_p = a_p^R + ia_p^I$, we obtain from Eqs.~(\ref{eq:grouprel}), (\ref{eq:T}) and (\ref{eq:S}) the following relations 
\begin{eqnarray}\label{eq:eq}
\sum_\alpha \sum _i \left|U_{\alpha i}\right|^2 \cos\left( \phi_\alpha + \psi_i \right) = a_p^R~, \nonumber\\
\sum_\alpha \sum _i \left|U_{\alpha i}\right|^2 \sin\left( \phi_\alpha + \psi_i \right) =  a_p^I~. 
\end{eqnarray} 
They are generalization of equations obtained in~\cite{Hernandez:2012ra} for $n = 2$. The Eq.~(\ref{eq:eq}) provide relations between moduli squared of the all elements of $U_{\rm PMNS}$ for specific ``symmetry assignment''. In general, all the elements of PMNS matrix are involved in each equation in ({\ref{eq:eq}). However, for some specific values of the phases $\phi_\alpha$ and $\psi_i$, these equations can be reduced to relations between the elements of one row or column of the $U_{\rm PMNS}$. This is the case if the phases are $0$ and $\pi$; {\it i.e.}, if $n=2$ or $m=2$.

The two relations in Eq.~(\ref{eq:eq}) between the 4 independent parameters (three angles and CP phase) allow, {\it e.g.}, to predict two of them, once two others are fixed by experimental data. If one adds an additional transformation $S$ or $T$, two more relations with different values of $\phi_\alpha$ and $\psi_i$ will appear. If consistent, this can fix all the mixing parameters completely. 

In what follows we obtain the relations for various symmetry assignments and confront them with experimental results, thus identifying phenomenologically viable possibilities. The symmetry assignment consist of specifying ({\it i}) von Dyck group parameters $(n,m,p)$, ({\it ii}) the values of neutrino charges under the transformation $S$, $(l_1,l_2,l_3)$, ({\it iii}) the values of charged lepton charges under the transformation $T$, $(k_e,k_\mu,k_\tau)$, ({\it iv}) the value of $a_p$, for $p\geq4$. 

Recall that $n = 2$ is the only possibility for Majorana neutrinos and therefore the only common case to both Majorana and Dirac neutrinos, which has been explored in~\cite{Hernandez:2012ra}. In this case $\psi_i=0$ and $\pi$ and the relations in Eq.~(\ref{eq:eq}) are reduced to relations between elements of the \textit{columns} of $U_{\rm PMNS}$. These two relations and unitarity condition fix the column completely. The column number $i$ is determined by the neutrino mass state $\nu_i$ which is invariant under $S$, {\it i.e.} has $\psi_i=0$. All the relations obtained in~\cite{Hernandez:2012ra} for the Majorana neutrinos are valid also for the Dirac neutrinos in the case $n = 2$. However, model-building for the Dirac neutrinos can be different. 

In what follows we search for new relations between mixing parameters which are specific to the Dirac neutrinos.

\section{Relations between the mixing parameters for $G_\ell = \mathbf{Z}_2$}
\label{sec:cons}

In the case of $m = 2$ the only symmetry assignments for the charged leptons consistent with the condition $\sum_\alpha k_\alpha =  mq_k$ are $k_\alpha = (0,1,1)$, $(1,0,1)$ and $(1,1,0)$. They correspond to one of the phases $\phi_\gamma=0$  and two others equal $\phi_{\alpha\neq\gamma}= \pi$. Inserting this set of phases in Eq.~(\ref{eq:eq}) we obtain
\begin{eqnarray}\label{eq:eqm2}
\sum _i \left( 2 |U_{\gamma i}|^2 -1 \right) \cos \psi_i = a_p^R, \nonumber\\
\sum _i \left( 2|U_{\gamma i}|^2 -1 \right) \sin \psi_i = a_p^I~.
\end{eqnarray}
Thus, the symmetry group relation, with the unitarity condition, determine all elements of the $\gamma$-row of the mixing matrix. Recall that in (\ref{eq:eqm2}) the index $\gamma$ corresponds to the lepton invariant under transformation $T$: $k_\gamma = 0$.

The solution to Eq.~(\ref{eq:eqm2}) is 
\begin{eqnarray}\label{eq:m2}
|U_{\gamma 1}|^2 & = & \frac{a_p^R \cos\frac{\psi_1}{2} + \cos\frac{3\psi_1}{2} - a_p^I \sin\frac{\psi_1}{2}}{4\sin\frac{\psi_{21}}{2} \sin\frac{\psi_{31}}{2}}~, \nonumber\\
|U_{\gamma 2}|^2 & = & \frac{a_p^R \cos\frac{\psi_2}{2} + \cos\frac{3\psi_2}{2} - a_p^I \sin\frac{\psi_2}{2}}{4\sin\frac{\psi_{12}}{2} \sin\frac{\psi_{32}}{2}}~, \nonumber\\
|U_{\gamma 3}|^2 & = & \frac{a_p^R \cos\frac{\psi_3}{2} + \cos\frac{3\psi_3}{2} - a_p^I \sin\frac{\psi_3}{2}}{4\sin\frac{\psi_{13}}{2} \sin\frac{\psi_{23}}{2}}~, 
\end{eqnarray}
where $\psi_{ij}\equiv\psi_i-\psi_j$~\footnote{Eq.~(\ref{eq:m2}) differs from relations in~\cite{Hernandez:2012ra} by substitution: $\phi_e\to\psi_1$, $\phi_\mu\to\psi_2$ and $\phi_\tau\to\psi_3$.}. The solution is valid for any value of $n$ and $p$.

There are interesting properties of Eq.~(\ref{eq:m2}) which we will use in our further considerations. If $a_p$ is real ($a_p^I = 0$) and one of the phases $\psi_i=0$, {\it i.e.} the corresponding symmetry parameter $l_i = 0$, the equations in Eq.~(\ref{eq:m2}) are invariant with respect to the permutation of two other (nonzero) parameters $l_j \leftrightarrow l_k$. So, the assignments $(0, l_2, l_3)$ and $(0, l_3, l_2)$ or $(l_1, l_2, 0)$ and $(l_2, l_1,0)$, {\it etc.} will lead to the same solution in Eq.~(\ref{eq:m2}). Indeed, $l_2 + l_3 = n$, or $\psi_2 = 2\pi-\psi_3$, therefore permutation of $l_2$ and $l_3$ is equivalent to: $\psi_2 \rightarrow  2\pi- \psi_2$, $\psi_3 \rightarrow 2\pi-\psi_3$. This change the sign of both numerator and denominator, thus leaving the whole expressions in Eq.~(\ref{eq:m2}) invariant.

We performed scan of all possible symmetry assignments, satisfying Eq.~(\ref{eq:vondyck}), and identified the phenomenologically viable ones; {\it i.e.}, symmetry assignments which lead to relations among the mixing parameters in agreement with experimental data. The latter has been done in the following way: for fixed values of $\theta_{13}$ and $\theta_{12}$, the two symmetry relations in Eq.~(\ref{eq:eqm2}) determine the values of $\theta_{23}$ and $\delta$. We find regions of $\theta_{23}$ and $\delta$ by varying the angles $\theta_{13}$ and $\theta_{12}$ in the $3\sigma$ allowed ranges from the global fit of the neutrino oscillation data. Then we confront this prediction with the allowed regions in the $(\sin^2\theta_{23},\delta)$ plane. The results are shown in Figure~\ref{fig:allowed} where the solid, dashed and dot-dashed curves show the allowed regions respectively at $1\sigma$, $2\sigma$ and $3\sigma$ confidence level from global analysis of neutrino oscillation data (including the latest T2K~\cite{Abe:2015awa} and NO$\nu$A~\cite{nova} data) taken from~\cite{marrone}. The black point inside the $1\sigma$ regions shows the best-fit point.

We find only three symmetry assignments that lead to predictions compatible with the experimental data, at least at $3\sigma$ level. These are shown by colored regions in Figure~\ref{fig:allowed}. 

All of these possibilities have one $l_j = 0$, and the corresponding phase $\psi_j = 0$. In this case the condition in Eq.~(\ref{eq:det1}) is reduced to $ l_i + l_k = n q_l$, $i, k \neq j$, or $\psi_i = - \psi_k + 2 \pi q_l$. We will consider general case of all nonzero values of $l_j$ in sec.~\ref{sec:m3}.

Below we describe the viable possibilities in detail.

1). $(n,m,p)=(3,2,5)$: the neutrino symmetry is $\mathbf{Z}_3$, with the embedding von Dyck group $\mathbf{A}_5$. Neutrino charges equal $(l_1,l_2,l_3)=(0,1,2)$ or $(0,2,1)$ which give the same result, and the trace parameter is $a_5=(1-\sqrt{5})/2$. Then for the charged lepton phase $\phi_\mu = 0$ ($\mu$-row), which we call the $\mu$-solution, or $\phi_\tau = 0$ ($\tau$-row, the $\tau$-solution) we obtain
\begin{equation}\label{eq:n3m2}
\left( |U_{\mu(\tau)1}|^2,|U_{\mu(\tau)2}|^2,|U_{\mu(\tau)3}|^2 \right) = \left(\frac{3-\sqrt{5}}{6},
\frac{3+\sqrt{5}}{12},\frac{3+\sqrt{5}}{12}\right)~.
\end{equation}
The case $\phi_e=0$ is inconsistent with the data. The main feature of solutions in Eq.~(\ref{eq:n3m2}) is that 
\begin{equation}
|U_{\mu(\tau)2}|^2 = |U_{\mu(\tau)3}|^2 \approx 0.436~. 
\end{equation}
In terms of the mixing angles the $\mu$-solution (in the standard parametrization of the PMNS matrix) reads
\begin{eqnarray}\label{eq:n3m2mu}
|U_{\mu 3}|^2 & = & s_{23}^2 c_{13}^2 = A~, \nonumber\\
|U_{\mu 2}|^2 & = & c_{12}^2c_{23}^2 + s_{12}^2s_{23}^2s_{13}^2 - 2 c_{12}c_{23}s_{12}s_{23}s_{13}\cos\delta = B, 
\end{eqnarray}
where $c_{ij}=\cos\theta_{ij}$, $s_{ij}=\sin\theta_{ij}$ and $A=B=(3+\sqrt{5})/12\approx0.436$. From the first equality in Eq.~(\ref{eq:n3m2mu}) we find that 
\begin{equation}
s_{23}^2 = \frac{A}{c_{13}^2} \approx 0.445~. 
\end{equation}
Due to high accuracy of the measurements of $\theta_{13}$ angle, and the smallness of this angle, the value of $\theta_{23}$ is fixed rather precisely. The second relation in Eq.~(\ref{eq:n3m2mu}) gives
\begin{equation}\label{eq:delmu}
\cos\delta = 2\;\frac{-B+c_{12}^2c_{23}^2+s_{13}^2s_{12}^2s_{23}^2}{\sin2\theta_{12}\,\sin2\theta_{23}\,\sin\theta_{13}}~,
\end{equation}
where $B=0.436$. The first two terms in the numerator of Eq.~(\ref{eq:delmu}) have comparable values and the last term, proportional to $s_{13}^2$, is small. Consequently, even small variation of $c_{12}^2$ lead to significant change in the value of $\cos\delta$. These features can be seen in Figure~\ref{fig:allowed}: the red region shows the predicted ranges of parameters $(\sin^2\theta_{23},\delta)$ which are obtained using Eq.~(\ref{eq:n3m2mu}) with uncertainties of $\theta_{13}$ and $\theta_{12}$ taken into account ({\it i.e.}, varying the $\theta_{12}$ and $\theta_{13}$ in $3\sigma$ allowed ranges). The black cross in the red region shows the prediction from Eq.~(\ref{eq:n3m2mu}) assuming the best-fit values of $\theta_{13}$ and $\theta_{12}$. 


\begin{figure}[t]
\centering
\subfloat[]{
\includegraphics[width=0.5\textwidth]{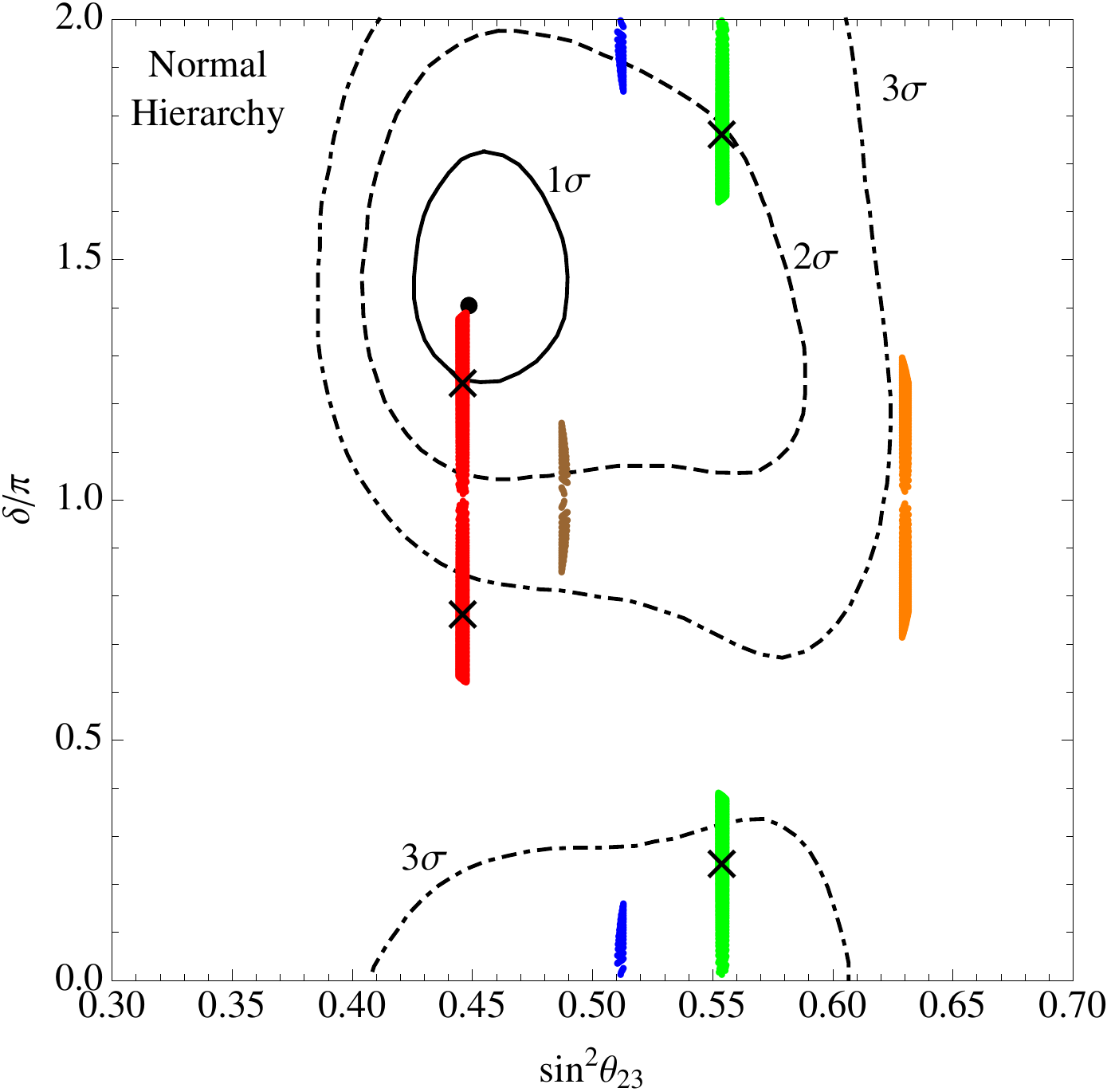}
\label{fig:nh}
}
\subfloat[]{
\includegraphics[width=0.5\textwidth]{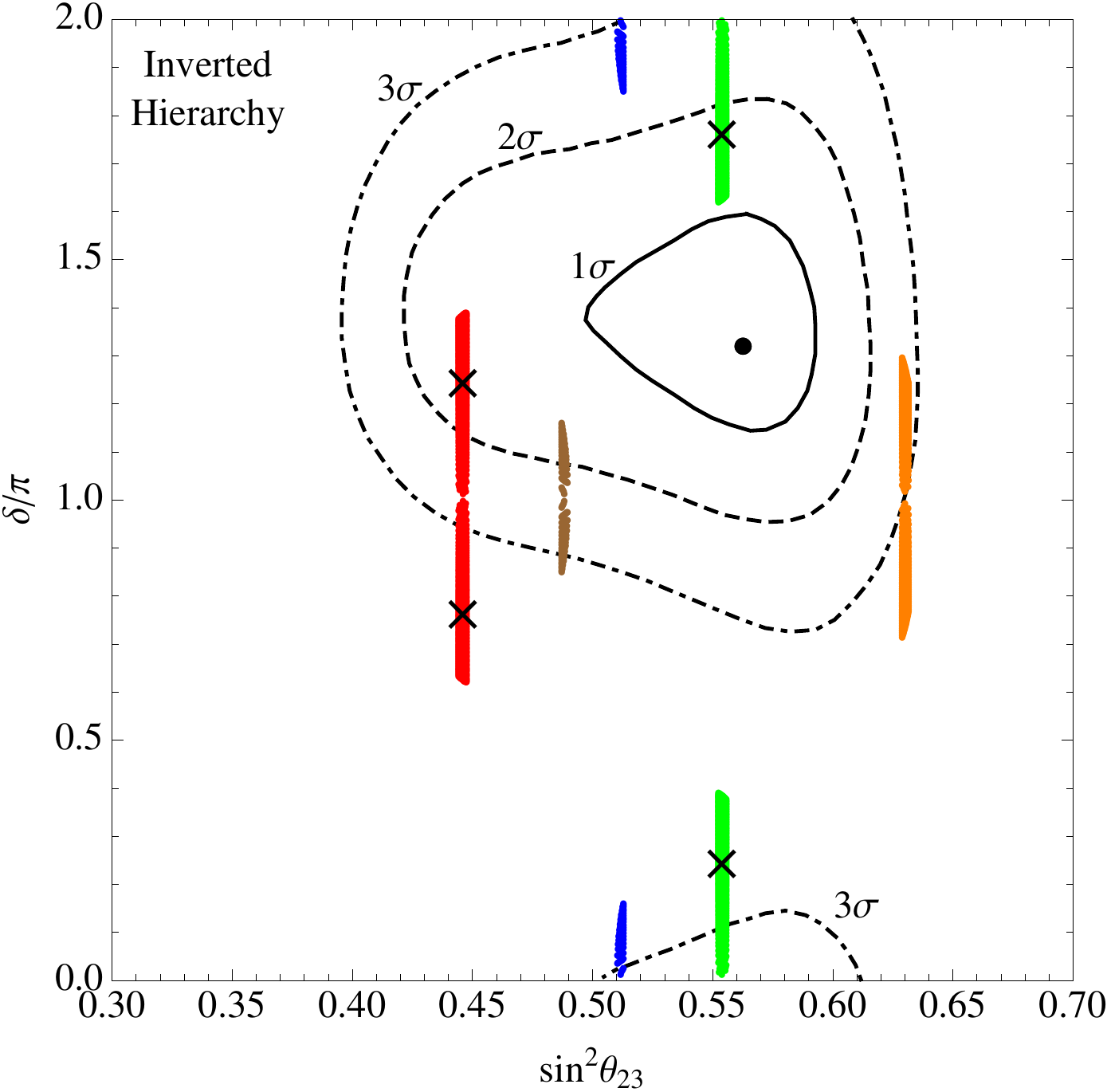}
\label{fig:ih}
}
\caption{\label{fig:allowed}The prediction of discrete symmetries for Dirac neutrinos in the $(\sin^2\theta_{23},\delta)$ plane, compared with the experimentally allowed region for normal (left panel) and inverted (right panel) hierarchies. The red and green regions correspond to the predictions of the symmetry group $D(3,2,5)$. The blue and brown regions are for $D(4,2,3)$; and the orange region is for $D(5,2,3)$. The solid, dashed and dot-dashed curves determine the allowed regions from global analysis of oscillation data, respectively at $1\sigma$, $2\sigma$ and $3\sigma$~\cite{marrone}.}
\end{figure}

The $\mu$-solution is in good agreement with data in the first quadrant in the case of normal mass hierarchy. The best fit value of the phase $\delta = 1.25 \pi$, and maximal allowed phase, $\delta = 1.4\pi$, is close to the present best experimental fit. The predictions have overlap with the allowed region at $\sim1\sigma$.

For the $\tau$-solution we have
\begin{eqnarray}\label{eq:n3m2tau}
|U_{\tau 3}|^2 & = & c_{23}^2 c_{13}^2 = A~, \nonumber \\
|U_{\tau 2}|^2 & = & c_{12}^2s_{23}^2 + s_{12}^2c_{23}^2s_{13}^2 + 2 c_{12}c_{23}s_{12}s_{23}s_{13}\cos\delta = B~,
\end{eqnarray}
where $A=B=(3+\sqrt{5})/12$. From the first equality we obtain $s_{23}^2 = 0.555$. This solution is related to the $\mu$-solution by $s_{23}^2 \leftrightarrow c_{23}^2$ and changing the sign of $\cos\delta$. This connection is reflected in Figure~\ref{fig:allowed} where the green region shows the predicted values in $\tau$-solution. Since for the $\tau$-solution $s_{23}^2 \leftrightarrow c_{23}^2$, numerically we get the same estimation for $\delta$ up to the change $\delta\to\pi-\delta$, which can be seen in Figure~\ref{fig:allowed}. The black cross in the green region shows the prediction from Eq.~(\ref{eq:n3m2tau}) for the best-fit values of $\theta_{13}$ and $\theta_{12}$.

The $\tau$-solution gives good agreement with data for inverted mass hierarchy. The predicted best fit value of phase is $\delta=1.76\pi$, and the lower bound $\delta\geq1.62\pi$. The solution overlaps with $\sim2\sigma$ allowed region.

Future determination of the octant of $\theta_{23}$ can disentangle $\mu$- and $\tau$-solutions.

2). $(n,m,p)=(4,2,3)$: the neutrino residual symmetry is $\mathbf{Z}_4$ and the von Dyck group is $\mathbf{S}_4$. The symmetry assignments $(l_1,l_2,l_3)=(1,3,0)$ or $(3,1,0)$, $\phi_\mu = 0$ ($\phi_\tau =0$) and $a_3=0$ leads to
\begin{equation}\label{eq:n4m2}
\left( |U_{\mu(\tau)1}|^2,|U_{\mu(\tau)2}|^2,|U_{\mu(\tau)3}|^2 \right) = \left( \frac{1}{4},\frac{1}{4},\frac{1}{2} \right)~.
\end{equation} 
This solution gives relations between mixing parameters as in Eqs.~(\ref{eq:n3m2mu}) and (\ref{eq:n3m2tau}), with $A=1/2$ and $B=1/4$. The predictions are shown by blue and brown regions in Figure~\ref{fig:allowed}, respectively for the $\mu$-row and $\tau$-row solutions. For the best fit values of $\theta_{13}$ and $\theta_{12}$ the relations in Eq.~(\ref{eq:n4m2}) cannot be satisfied. Thus, there are no cross points in these regions. The 2-3 mixing is close to maximal: $s_{23}^2=0.5/c_{13}^2\approx0.511$ for the $\mu$-solution; and $s_{23}^2\approx0.489$ for the $\tau$-solution. Expression for phase, given by Eq.~(\ref{eq:delmu}) with $a=0.25$, leads to values of $\delta$ close to $0$ or $\pi$. The $\delta-\theta_{23}$ predictions are compatible with the data at best at $2\sigma$ level.

3) $(n,m,p)=(5,2,3)$: in this case the neutrino residual symmetry is $\mathbf{Z}_4$ and the covering von Dyck group is $\mathbf{A}_5$. The symmetry assignments $(l_1,l_2,l_3)=(0,3,2)$ or $(0,2,3)$, $\phi_\tau = 0$ and $a_3=0$ lead to
\begin{equation}\label{eq:n5m2}
\left( |U_{\tau1}|^2,|U_{\tau2}|^2,|U_{\tau3}|^2 \right) = \left( \frac{2}{5+\sqrt{5}},\frac{1+\sqrt{5}}{4\sqrt{5}},\frac{1+\sqrt{5}}{4\sqrt{5}} \right)~.
\end{equation}
Rewriting this relation in terms of mixing parameters gives relations similar to Eq.~(\ref{eq:n3m2tau}) with $A=B=(1+\sqrt{5})/(4\sqrt{5})\approx0.36$. The prediction from Eq.~(\ref{eq:n5m2}) is shown by the orange region in Figure~\ref{fig:allowed}. According to Eq.~(\ref{eq:n5m2}) 
\begin{equation}
|U_{\tau 2}|^2 = |U_{\tau 3}|^2 =A \approx 0.36~, 
\end{equation}
which deviates from maximal mixing significantly being at the border of $3\sigma$ allowed region. Again $\delta$ is close to $\pi$. The $\mu$-solution with $\phi_\mu=0$ is excluded at $3\sigma$ level.

The symmetry of the prediction regions in Figure~\ref{fig:allowed} with respect to $\delta=\pi$ comes from the $\cos\delta$ dependence of the moduli of mixing elements.

\section{The cases with $G_\ell=\mathbf{Z}_m$ and $m>2$}
\label{sec:m>2}

As we have mentioned in sec.~\ref{sec:residual}, for symmetry groups with $n,m>2$ the relations in Eq.~(\ref{eq:eq}) cannot be reduced to constraints on the elements of single column or row of the $U_{\rm PMNS}$ matrix. We find that all the symmetry assignments with $m>2$ (which include $D(3,3,2)$, $D(3,4,2)$, $D(3,5,2)$, $D(4,3,2)$ and $D(5,3,2)$) lead to relations between matrix elements which are incompatible with the experimental data at $3\sigma$ level. To elucidate the reason behind this result we consider one example in details.  

$D(3,3,2)\equiv \mathbf{A}_4$: for this assignment the first relation in Eq.~(\ref{eq:eq}) leads to
\begin{equation}\label{eq:n3m3first}
|U_{\alpha i}|^2 + |U_{\beta j}|^2 + |U_{\gamma k}|^2 = \frac{1}{3}~,
\end{equation}
where $\alpha$, $\beta$ and $\gamma$ are all different and should be selected from $(e,\mu,\tau)$, also $i$, $j$ and $k$ are all different being selected from $(1,2,3)$ depending on various assignments for $(k_e,k_\mu,k_\tau)$ and $(l_1,l_2,l_3)$. 

The key feature of the equality in Eq.~(\ref{eq:n3m3first}) is smallness of the right side which strongly restricts the allowed elements in the left side. Since the elements $|U_{e1}|^2$, $|U_{\mu 3}|^2$ and $|U_{\tau 3}|^2$ alone are bigger than $1/3$, the left hand side should include $|U_{e3}|^2$ and small elements from the $\mu$ and $\tau$ rows and the 1 and 2 columns. In fact, the only allowed choices of combinations of elements in Eq.~(\ref{eq:n3m3first}) are $|U_{e3}|^2 + |U_{\mu1}|^2 + |U_{\tau2}|^2 =1/3$ and a similar combination with $\mu\leftrightarrow\tau$. The second relation in Eq.~(\ref{eq:eq}) for the first combination leads to the equality
\begin{equation}\label{eq:n3m3second1}
|U_{e1}|^2 + |U_{\mu 3}|^2 + |U_{\tau 2}|^2 = |U_{e2}|^2 + |U_{\mu 1}|^2 + |U_{\tau 3}|^2~,
\end{equation}
which can be reduced to 
\begin{equation}
c_{13}^2 \left[ \cos2\theta_{12}-\cos2\theta_{23} \right] =0~,
\end{equation}
and $\theta_{23}=\theta_{12}$. For the second combination the relation can be obtained by $\mu\leftrightarrow\tau$ permutation, which finally leads to $\theta_{23}=\pi/2-\theta_{12}$. Both relations are excluded by experimental data. Notice, however, that the current best fit values are $\sin\theta_{23}=0.67$ and $\sin\theta_{12}=0.56$, so that $\sim10\%$ corrections to the equality $\sin\theta_{23}=\sin\theta_{12}$ can bring these relations to agreement with the data.

For the other symmetry groups with $m>2$ similar arguments can be found showing the exclusion by experimental data, which we have checked also by numerical calculation. 

\section{Further considerations and generalizations}
\label{sec:m3}

Here we lift some of the conditions imposed in the previous sections. 

1. Let us relax condition that one of the phases $\psi_i$ is zero. Notice that for $m = 2$ the equality $\phi_\alpha = 0$ for one the phases is unavoidable. It is straightforward to check that the equality $\det[S]=1$ can be satisfied with all nonzero values of $l_i$ only for $n\geq3$. In the following we take $m=2$ and explore the cases with $n\geq3$.

For $n\leq5$, the condition $\det[S]=1$ or (\ref{eq:det1}) can be satisfied (with all the $l_i$ being nonzero) when at least two of the $l_i$ are equal. Indeed, for $n=3$, the possibilities are $l_j = (1, 1, 1)$ ($q_l = 1$) and $l_j = (2, 2, 2)$ ($q_l = 2$). In the case of all equal charges $l_j$, and therefore the phases $\psi_j = \psi$, the conditions in Eq.~(\ref{eq:eqm2}) are reduced (due to unitarity) to the inconsistent equalities $\sin \psi = \cos \psi = 0$. For $n=4$ two assignments satisfy condition (\ref{eq:det1}): $l_j =(1, 1, 2)$ ($q_l = 1$) and $l_j =(2,3,3)$ ($q_l = 2$). For $n= 5$ we have $(1,1,3)$, $(1,2,2)$ ($q_l = 1$) and $(4,3,3)$, $(4,4,2)$ ($q_l = 2$) (with all possible permutations of charges). General feature of all these assignments is that two charges and therefore two phases are equal. Denoting equal phases by $\psi^\prime$, we get for the remaining phase $\psi_j=2\pi q_l - 2\psi^\prime$. Then the relations in Eq.~(\ref{eq:eqm2}) are reduced to
\begin{equation}\label{eq:4-5rel}
\left| U_{\gamma j} \right|^2 =\frac{a_p^R+\cos2\psi^\prime}{2\left( \cos2\psi^\prime 
- \cos\psi^\prime \right)} \quad,  ~~~~~ \quad \left| U_{\gamma j} \right|^2 
=\frac{-a_p^I+\sin2\psi^\prime}{2\left( \sin2\psi^\prime + \sin\psi^\prime \right)}~,
\end{equation}
where $\gamma$ corresponds the charge lepton with zero phase $\phi_\gamma=0$ and $j$ --- to the neutrino with unequal phase. 

Explicitly, for $n = 4$ we have $\psi^\prime= \pi/2$ or $\psi^\prime= 3\pi/2$ and then Eq.~(\ref{eq:4-5rel}) gives $\left| U_{\gamma j} \right|^2 = 0.5 (1 - a_p^R)$ and $\left| U_{\gamma j} \right|^2 = - 0.5 a_p^I$. These two equalities are inconsistent for $a_2$ and $a_3$. For $p=4$ the finiteness condition (\ref{eq:vondyck}) is not satisfied. Assuming that in this case a subgroup can be made finite and using $a_4=1$ or $-1-2i$ we obtain $|U_{\gamma j}|^2=0$ or $1$. The first case may still work for $\gamma = e$ and $j = 3$, \textit{i.e.} for $U_{e3}$, in the first approximation.

For $n = 5$, the phase $\psi^\prime$ can be $2\pi/5$, $4\pi/5$, $6\pi/5$ and $8\pi/5$. We have checked that none of these $\psi^\prime$ values, and possible values of $a_p$, lead to a consistent solution of Eq.~(\ref{eq:4-5rel}). 

For $n\geq6$, there are more assignments that satisfy the condition (\ref{eq:det1}) with all nonzero $l_j$, both when two $l_j$ are equal and when they are all different. Furthermore, the finiteness of group, Eq.~(\ref{eq:vondyck}), requires that $p=2$ and consequently $a_2=-1$. When two $l_j$ are equal, the relations in Eq.~(\ref{eq:eqm2}) lead to $\cos2\psi^\prime=\cos\psi^\prime$ and therefore to zeros in the denominator of relations in Eq.~(\ref{eq:4-5rel}). For the case of all different $l_j$, we have numerically checked that no viable solution of Eq.~(\ref{eq:eqm2}) exist.

Thus, generally, when $m=2$, there is no viable symmetry assignment with all nonzero charges $l_j$. \\

2. We have numerically checked that also for $m>2$ no symmetry group relation compatible with the data and with all nonzero charges $l_j$ can be obtained. \\

3. In all the considered examples in secs.~\ref{sec:cons} and \ref{sec:m>2} symmetry fixes two out of four mixing parameters in $U_{\rm PMNS}$. All the 4 mixing parameters can be determined by symmetry if additional symmetry transformation, $S^\prime$ or $T^\prime$, is introduced. This gives second symmetry group relation which can lead to two new relations between the matrix elements. 

Let us consider the second transformation $T^\prime$. We can take $m^\prime=2$ (but with different assignment for $k_\alpha$ than the assignment for $T$). In this way we can fix the elements of both $\mu$- and $\tau$-rows. However, as we see in Figure~\ref{fig:allowed} the $\mu$- and $\tau$-solutions are inconsistent with each other (the corresponding regions do not overlap). In sec.~\ref{sec:m>2} we have shown that for $m^\prime>2$ there is no solution compatible with data for any $n\geq3$.

Let us consider a second neutrino transformation $S^\prime$ and single $T$ with $m=2$. For $n^\prime\geq3$ again the second symmetry group condition will give one of the viable solutions described above. Again, since any two different obtained solutions are incompatible, adding such $S^\prime$ will be inconsistent.

Finally let us consider the case of $S^\prime$ with $n=2$ (and $T$ with $m=2$). In this case the second symmetry group condition is characterized by $n=m=2$ and arbitrary $p$. In this case the phases of transformations are $0$ or $\pi$, both in  $S^{\prime}$ and $T$. Therefore, the left-hand side of the second equation in (\ref{eq:eq}) is zero, which requires $a_p^I=0$, for consistency. So, in this case only the first equation in (\ref{eq:eq}) gives non-trivial relation on the mixing elements. From this equation we obtain
\begin{equation}\label{eq:nn}
|U_{\alpha i}|^2 = \frac{1+ a^R_p}{4}~,
\end{equation}
where $\alpha$ and $i$ are referred to the charged lepton and neutrino which are invariant under transformations. In this way, in principle, one can fix any element of the mixing matrix. Possible values of the right-hand side in Eq.~(\ref{eq:nn}) are 0 (for $a_2$), $1/4$ (for $a_3$), 1/2 (for $a_4$), and $0.65$ or $0.08$ (for $a_5$). The interesting possibility would be $|U_{\alpha i}|^2 = 0.65$ for $a_5$, which is close experimental result for $|U_{e1}|^2$. However, in the case of single $T$, it fixes $\alpha$ for the additional relation, which should be $\mu$ or $\tau$. So, it can only fix one of the elements of the row which was already determined by $S$ and $T$. To fix the element outside the fixed row by the first symmetry group relation one needs to introduce another transformation $T'$, and therefore further expand the symmetry group.

\section{Conclusion}
\label{sec:conc}

We have studied mixing of the Dirac neutrinos in the residual symmetries approach. The key difference from the Majorana case (extensively studied before) is that the Dirac mass matrix may have larger generic (mass independent) symmetries: $\mathbf{Z}_{n}$  with $n > 2$, with maximal symmetry being the product of three such factors. For the Majorana neutrinos only $n = 2$ is possible with the maximal symmetry $\mathbf{Z}_{2} \times \mathbf{Z}_{2}$. Of course for the Dirac neutrinos also the case $n = 2$ is applied.

We generalized the symmetry group relations to the case of Dirac neutrinos and use them to explore new patterns of lepton mixing which can be obtained for the Dirac case. The residual neutrino symmetries $\mathbf{Z}_{n}$ with $n \geq 3$ have been explored. 

We have found all new phenomenologically viable (within allowed $3\sigma$ region) relations between mixing parameters and the corresponding symmetry assignments. We presented the relations as predictions for the 2-3 mixing angle, $\theta_{23}$, and the CP phase, $\delta$, in terms of the well-measured parameters $\sin^2 \theta_{13}$ and $\sin^2 \theta_{12}$. 

We find that for $n \geq 3$ the viable solutions (relations) exist only if the charged lepton residual symmetry is $G_{\ell} = \mathbf{Z}_{2}$. In this case the symmetry fixes elements of a single row of the PMNS matrix. Viable solutions (for mixing angles) exist for $n = 3$ and the group $\mathbf{A}_5$, $n = 4$ and the group $\mathbf{S}_4$, $n = 5$ and the group $\mathbf{A}_5$. The solutions fix rather precisely $\theta_{23}$ and give rather large range of values of $\delta$ which is determined by the present uncertainty in $\theta_{12}$.   

We found that introduction of the second $S^\prime$ or $T^\prime$, which can fix all the mixing parameters, does not lead to consistent solution.   

For bigger residual symmetry of the charged leptons, $G_{\ell} = \mathbf{Z}_{m}$ with $m > 2$, the relations between the mixing matrix elements become more complicated, involving elements of both columns and rows. We checked that no viable solutions exist in this case. At the same time some interesting relations are realized which can be brought in agreement with data if some relatively small corrections (related to violation of the residual symmetries) are included. In particular, the equality $\theta_{23} = \theta_{12}$ is realized in $D(3, 3, 2) = \mathbf{A}_{4}$ case. Corrections of the order $10\%$ can lead to agreement with data. 

The solutions we have found lead to discrete values of $\sin^2\theta_{23}$, determined by the symmetry group parameters. Future precise measurement of $\sin^2\theta_{23}$ with accuracy $\sim0.01$ can discriminate among the possibilities. Also, the precise determinations of $\sin^2\theta_{12}$ ({\it e.g.} by JUNO) will lead to precise prediction of the phase $\delta$, and so future measurements of the phase will provide crucial checks of the obtained symmetry relations.    

On the other hand, the identified viable symmetries and symmetry assignments will be useful for further model building.


\begin{acknowledgements}
The authors would like to express special thanks to the Mainz Institute for Theoretical Physics (MITP) for its hospitality and support. A.~E. thanks the Max-Planck-Institut f\"ur Kernphysik, Heidelberg, for the hospitality and support during the completion of this work.
\end{acknowledgements}


\end{document}